\documentclass[aps,showpacs,preprint,preprintnumbers,amssymb,amsmath,amsfonts,groupedaddress]{revtex4}

\usepackage{graphicx}
\usepackage{dcolumn}
\usepackage{bm}

\newcommand{\bmath}{\begin{mathletters}}
\newcommand{\emath}{\end{mathletters}}
\newcommand{\be}{\begin{eqnarray}}
\newcommand{\ee}{\end{eqnarray}}
\newcommand{\ba}{\begin{array}}
\newcommand{\ea}{\end{array}}

\newcommand{\no}{\nonumber}

\newcommand{\ga}{\gamma}

\newcommand{\rtar}{\rightarrow}

\newcommand{\txtc}  {{\mathrm{c}}}

\newcommand{\txts}  {{\mathrm{s}}}
\newcommand{\txtd}  {{\mathrm{d}}}
\newcommand{\txtm}  {{\mathrm{m}}}

\newcommand{\txtL}  {{\mathrm{L}}}
\newcommand{\txtS}  {{\mathrm{S}}}
\newcommand{\txtY}  {{\mathrm{Y}}}

\newcommand{\vecn} {{\vec{n}}}

\newcommand{\vecr} {{\vec{r}}}

\begin{document}

\title{Ground-State Shapes and Structures of Colloidal Domains}

\author{Jianlan Wu and Jianshu Cao\footnote{Email: jianshu@mit.edu}}
\affiliation{Department of Chemistry, Massachusetts Institute of Technology \\ Cambridge, MA 02139}

\date{First submitted August 21, 2004}

\begin{abstract}

In charged colloidal suspensions, the competition
between  square-well attraction and long-range Yukawa repulsion
leads to various stable domains and Wigner supercrystals.
Using a  continuum model and symmetry arguments,
a phase diagram of  spheres, cylinders, and lamellae is obtained as a function of two
control parameters, the volume fraction and the ratio between the surface tension
and repulsion. Above a critical value of the ratio, the microphase
cannot be supported by the Yukawa repulsion and macroscopic phase separation occurs.
This finding quantitatively explains the lack of pattern formation
in simple liquids because
of the small hard sphere diameter in comparison with the size of macromolecules.
The  phase diagram also predicts microphase separation at zero value of the ratio,
suggesting the possibility of self-assembly in repulsive systems.

\end{abstract}
\pacs{82.70.Dd, 05.65.+b, 64.70.-p}

\maketitle

\newpage

\begin{figure}[h]
  \includegraphics[width=5cm]{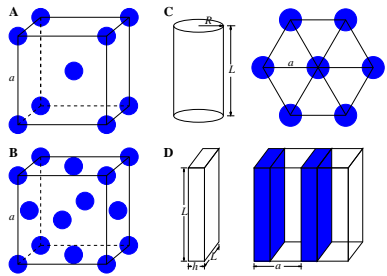}
  \caption{Super structures of domains: A) bcc lattice of spherical domains,
        B) fcc lattice of spherical domains, C) a cylindrical domain and
        the resulting 2D triangular lattice, D) a lamellar domain and
        the resulting 1D lattice.}
    \label{fig00}
\end{figure}

Microphase separation is ubiquitous in soft matter
systems\cite{seul95,leibler,mcconnell01,blankschtein,gelbart,lipowsky,widom,wolynes00}.
For example, microphase separation in block copolymers
results from mixing of two or more chemically different polymer segments\cite{leibler}.
Competition of hydrophobic and hydrophilic head-groups of amphiphiles  leads to
micelles in the water solution\cite{blankschtein}.
The self-assembly processes in charged colloidal suspensions
and protein solutions can lead to
the formation of stable domains and Wigner supercrystals (see Fig.~\ref{fig00})
at low temperatures.
The investigation of charged colloids and protein solutions is particularly
interesting because they can be approximated as one-component systems
with an effective isotropic pairwise interaction after averaging
out the degrees of freedom of dispersing medium.
This effective interaction is composed of a hard-core potential,
a long-range electrostatic repulsion, and a short-range attraction.
The screened Coulomb repulsion is described by a Yukawa potential\cite{Verwey,Wu}.
Since more complicated potential forms qualitatively lead to the
same phenomena\cite{sciortino,caowu042},
for simplicity, we model the short-range attraction by a
square-well potential.
The overall pairwise interaction between two colloidal particles
separated by $r$ is given by
\be
u(r) = \left\{\ba{lll} \infty &~~~~ r\le \sigma  \\ -\varepsilon &~~~~ \sigma<r\le \lambda\sigma
\\ u_{\txtY}(r)=A\zeta r^{-1} e^{-r/\zeta} &~~~~ r>\lambda\sigma \\ \ea \right. ,
\label{eq00_01}
\ee
where $\sigma$ is the colloidal diameter.
The attraction depth $\varepsilon$ and repulsion strength $A$
are temperature dependent, and $\varepsilon$
is usually greater than the average thermal fluctuation.
For convenience all the length variables in this letter are
dimensionless in units of the screening length $\zeta$.
The pairwise and isotropic interaction in Eq.~(\ref{eq00_01}) represents
one of the simplest self-assembly systems that can be studied explicitly.

At low temperatures,  shapes and sizes of colloidal domains are narrowly distributed, and
their arrangements in space are highly ordered.
A perfect superlattice composed of identical colloidal domains
is an ideal reference (ground) state. In this letter we
concentrate on domain patterns in the
ground state. Entropic effects induced by thermal
fluctuations will be studied elsewhere.
Without thermal effects, we determine the most stable~(optimal) shape, size,
and superlattice (see Fig.~\ref{fig00}) for ground-state
domain patterns by globally minimizing
the  energy density.
At low temperatures, this evaluation is
simplified by a continuum approximation:
Colloidal particles are closely packed inside a domain
and the characteristic domain size is much larger than the particle size.
Similar models have been used in the study of
two-dimensional (2D) lipid domains\cite{mcconnell01}.
In our continuum model, the short-range attraction gives rise
to bulk adhesion and surface energy.
The density of the adhesive energy is a constant in the leading order
for a given colloidal number density $\rho$ or equivalently a given
volume fraction, $\phi=\pi\rho\sigma^3/6$. The bulk adhesion does not
affect the domain shapes and is not included in this letter.
For a domain with area $S$, the surface energy is
$U_{\txtS} = \ga S$, where the surface tension $\ga$ is proportional to the attraction
depth $\epsilon$ in the lowest order approximation.
Throughout this letter, $\ga$ is assumed to be independent of domain shapes.
The sum of the long-range Yukawa potential is separated into two parts,
\be
\sum_{i<j} u_{\txtY}(r_{ij})
=&& \sum_{m=1}^{N_\txtd} U^{(1)}_{\txtY}(m)+\sum_{m<n}^{N_\txtd} U^{(2)}_{\txtY}(m, n) \no \\
=&& \sum_{m=1}^{N_\txtd}\frac{\rho^2_1}{2}\int_{v_{\txtd, m}} {\mathrm{d}}\vecr_1
\int_{v_{\txtd, m}} {\mathrm{d}}\vecr_2 u_{\txtY}(r_{12}) \no \\
+&&\sum_{m<n}^{N_\txtd}\rho^2_1 \int_{v_{\txtd, m}} {\mathrm{d}}\vecr_1\int_{v_{\txtd, n}} {\mathrm{d}}\vecr_2 u_{\txtY}(r_{12}) ,
\label{eq00_02}
\ee
where $U^{(1)}_\txtY(m)$ is the intra-domain repulsion for domain $m$,
$U^{(2)}_\txtY(m, n)$ is the inter-domain repulsion between domains $m$ and $n$,
and $N_\txtd$ is the total number of domains.
Here $\rho_1$ is the colloidal number density
within domains and larger than the overall number density $\rho$.
The self energy of a single domain is the sum of
the intra-domain repulsion and the surface energy, $E_1 = U_\txtS+U^{(1)}_\txtY$.
The sum of the inter-domain repulsions
results in the lattice energy, $U_\txtL  = (1/2)\sum_{n(\neq 1)}^{N_\txtd} U^{(2)}_\txtY(1, n)$.
In the ground state, all the domains in one phase
are identical so that the energy of each domain
is given by $E_{\mathrm{tot}} = E_1+U_L$.
In our continuum model, the morphologies of domain
patterns are thus determined by minimizing $E_{\mathrm{tot}}/v_\txtd$,
where $v_\txtd$ is the domain volume.

\begin{figure}
  \includegraphics[width=6.5cm]{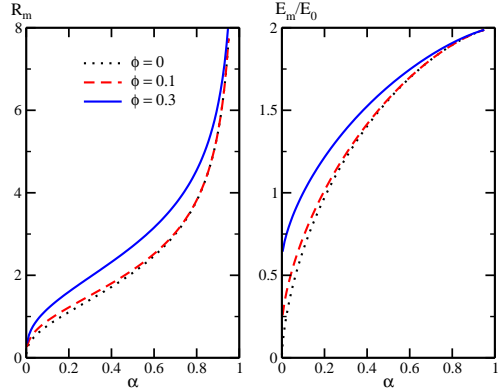}
  \caption{The optimal radius $R_\txtm$ and the energy minimum $E_{1, \txtm}$ ($E_{\mathrm{tot}, \txtm}$)
        of a spherical domain in the isolated case~(dotted lines) and in
       fcc lattices at finite volume fractions,
       where the dashed lines are for $\phi=0.1$ and the solid
        lines are for $\phi=0.3$.
    \label{fig01b}}
\end{figure}

We now apply the continuum model to study various domain patterns
shown in Fig.~\ref{fig00}. Spheres have the smallest surface  energy
and highest spatial symmetry. For a spherical domain with radius
$R$, the intra-domain Yukawa contribution $U^{(1)}_\txtY$ is
simplified using the Fourier transform technique. The summation of
the surface energy $U_\txtS$ and the intra-domain repulsion
$U^{(1)}_\txtY$ results in the self energy of a spherical domain,
\be E_1 =
\frac{E_0}{R^3}\left[2{R}^3+3(\alpha-1){R}^2+3-3(1+{R})^2e^{-2{R}}\right],
\label{eq01_07} \ee where $E_0=v_\txtd \varepsilon_0$ arises from a
constant energy density $\varepsilon_0=\pi\rho^2_1A\zeta^3$ and the
domain volume $v_\txtd = 4\pi R^3/3$. As shown in
Eq.~(\ref{eq01_07}), the competition between the surface tension and
the Yukawa repulsion is described by a single control parameter,
$\alpha=\gamma/\pi\rho^2_1A\zeta^4$. In the low density limit,
domains are far apart from one another so that an individual domain
can be treated as an isolated system and the lattice energy can be
neglected. The self energy has a minimum at a finite radius for
$\alpha<1$, whereas $E_1$ is a monotonously decreasing function of
$R$ for $\alpha\ge 1$. Thus $\alpha_\txtc=1$ is a critical point:
Spherical domains with finite sizes are stable for
$\alpha<\alpha_\txtc$, whereas the phase separation occurs for
$\alpha\ge\alpha_\txtc$. The self energy minimum $E_{1, \txtm}$ and
the associated radius $R_\txtm$ are plotted in Fig.~\ref{fig01b}. In
the limit of strong repulsion~($\alpha\rtar 0$), we obtain two
asymptotic forms, $ R_\txtm \sim \sqrt[3]{15\alpha}/2$ and $E_{1,
\txtm}/E_0 \sim 3\sqrt[3]{3\alpha^2/5}$. The size of the spherical
domain grows as $\alpha$ increases. In the phase separation limit
($\alpha\rtar\alpha^-_\txtc$), the stable radius and energy minimum
are asymptotically given by $R_\txtm \sim
\sqrt{3}\left(1-\alpha/\alpha_\txtc\right)^{-1/2}$  and $E_{1,
\txtm}/E_0\sim 2[1-(1-\alpha/\alpha_\txtc)^{3/2}/\sqrt{3}]$,
respectively. These asymptotic relations can be tested
experimentally.

\begin{figure}
  \includegraphics[width=6.5cm]{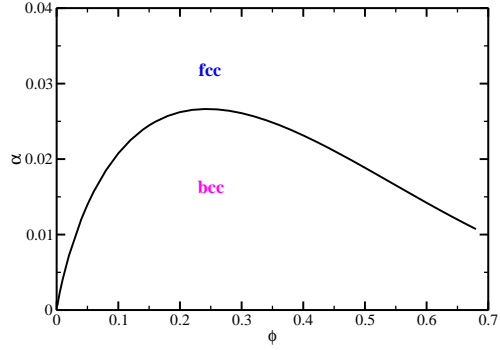}
  \caption{The phase diagram for fcc and bcc supercrystals composed by spherical domains.}
    \label{fig02}
\end{figure}

For a finite density (volume fraction), the balance of the self energy $E_1$
and the lattice energy $U_\txtL$ from inter-domain repulsions
leads to a 3D supercrystal, e.g., body centered cubic~(bcc)
and face centered cubic~(fcc) lattices of spheres (see Fig.~\ref{fig00}A and B).
The lattice energy $U_\txtL$ depends on the inter-domain repulsion
and the lattice structure.
Using the spherical harmonic expansion method,
the inter-domain repulsion between two spheres separated by $r$ is
derived as
\be
U^{(2)}_{\txtY}(r)
= \frac{3E_0}{{R}^{3}}\left[({R}+1)e^{-{R}}+(R-1)e^{{R}}\right]^2
\frac{e^{-{r}}}{r}.
\label{eq01_12}
\ee
The spatial periodicity of a Wigner lattice is related
to the volume fraction. For example, the lengths
of primitive cells for bcc and fcc lattices (see Fig.~\ref{fig00}A and B)
are given by $a = \eta \phi^{-1/3}R$, where we have
$\eta_{{\mathrm{fcc}}}=(16\pi/3)^{1/3}$ for the
fcc lattice, and $\eta_{{\mathrm{bcc}}}=(8\pi/3)^{1/3}$ for the bcc lattice.
Using the spatial periodicity, we sum the inter-domain repulsions
of the fcc (bcc) lattice and obtain the lattice energy as
\be
U_{\txtL} = && \frac{3E_0\phi^{1/3}}{\eta{R}^{4}}
\left[({R}+1)e^{-{R}}+(R-1)e^{{R}}\right]^2 \no \\
&&~~~~\times\sum_{\vecn\neq 0}
\frac{\exp(-\eta\phi^{-1/3}{R}x_\vecn/2)}{x_\vecn}.
\label{eq01_16}
\ee
The reduced distance $x_{\vecn}$ in the above equation is
given by $x_{\vecn}=[2(n^2_1+n^2_2+n^2_3+n_1n_2+n_2n_3+n_3n_1)]^{1/2}$ for the fcc lattice,
and $x_{\vecn}=[3(n^2_1+n^2_2+n^2_3)-2(n_1n_2+n_2n_3+n_3n_1)]^{1/2}$ for the bcc lattice,
where $n_i$ can be any integers except for that all the $n_i$ are zero.
The total  energy of a domain is the sum of the self energy in Eq.~(\ref{eq01_07})
and the lattice energy in Eq.~(\ref{eq01_16}).
Minimization of the energy density  for a given lattice
using $\partial_R (E_{{\mathrm{tot}}}/E_0)=0$
leads to the stable radius $R_\txtm$.
Figure~\ref{fig01b} shows that both $R_\txtm$
and the energy minimum $E_{{\mathrm{tot}}, \txtm}$
increase with the volume fraction.
By examining the subtle difference~($\sim 10^{-4}$)
of $E_{{\mathrm{tot}}, \txtm}$ between fcc and bcc lattices, we obtain the
phase diagram displayed in Fig.~\ref{fig02}. It demonstrates that the fcc lattice
is usually more stable than the bcc lattice
except for small values of $\alpha$.
Our results extend the previous studies on the
fcc-bcc transition in Wigner particle solids with the Yukawa potential\cite{fcc_bcc}.

Dispersing medium surrounded by colloidal particles
can be considered as cavities.
For $\phi>0.5$, cavities are dispersed in the sea of colloidal particles so that
various shapes and structures of cavities can form in the same way as domains.
The intrinsic mirror symmetry between conjugate colloidal domains and cavities
requires that the optimal shape and structure of cavities
at $\phi(>0.5)$ are the same as those of colloidal domains
at $1-\phi$. As a result, the cavity phase and the domain phase must
be equivalent at $\phi=0.5$, which can be achieved by the lamellar
shape and its topological variations. Following the same
domain-cavity symmetry argument, lamellae are expected to
be preferred in the phase separation limit ($\alpha\rtar\alpha^-_\txtc$).

Next we investigate lamellar domains.
For a lamellar domain with a finite width $h$, and infinite length and height $(L\rtar\infty)$,
its surface energy is given by $U_{\txts}= 2\gamma(L^2 +2Lh) \approx 2 E_0 \alpha {h}^{-1}$,
where $E_0=v_\txtd\varepsilon_0=L^2h\varepsilon_0$.
Following Eq.~(\ref{eq00_02}), we obtain the intra-domain repulsive energy,
\be
U^{(1)}_{\txtY} &&= 2 E_0 {h}^{-1}\left[h-1+e^{-h}\right].
\label{eq02_02}
\ee
In the ground state, all lamellar domains are parallel and form
a one-dimensional supercrystal (see Fig.~\ref{fig00}D).
To obtain the lattice energy, we calculate the repulsion  energy
between two parallel lamellar domains separated by $r$,
\be
U^{(2)}_\txtY (r) = 2 E_0 {h}^{-1} e^{-r} (e^{h}+e^{-h}-2).
\label{eq02_05}
\ee
At the volume fraction $\phi$, the distance between two arbitrary lamellar domains is
given by $|n|\phi^{-1} h$, where $n$ is a nonzero integer.
The lattice energy for each lamellar domain is  given by
\be
U_{\txtL} && = E_0 {h}^{-1}\left(e^{h}+e^{-h}-2\right)\sum_{n\neq 0} e^{-|n| \phi^{-1}h} \no \\
&& = 2 E_0 \frac{e^{h}+e^{-h}-2}{h(e^{h/\phi}-1)}.
\label{eq02_06}
\ee
We determine the optimal width $h_\txtm$
from minimization of $E_{\mathrm{tot}}/E_0$.
Similar to spherical domains, lamellar domains with a finite width can exist
only for $\alpha<\alpha_c$.
In the limit of strong repulsion~($\alpha\rtar 0$) at a finite volume fraction,
the lamellar width and the total energy are asymptotically given by
$h_\txtm \sim \sqrt[3]{6\phi\alpha/(1-\phi)^2}$ and
$E_{{\mathrm{tot}}, \txtm}/E_0 = 2\phi+\sqrt[3]{9(1-\phi)^2\alpha^2/(2\phi)}$, respectively.
By comparing $E_{{\mathrm{tot}}, \txtm}$ of spheres and lamellae,
we obtain their relative stability, as shown
in Fig.~\ref{fig04}. As expected, lamellar shapes are more stable than spheres
when $\alpha$ approaches the critical value $\alpha_\txtc$, or when
the volume fraction approaches $0.5$.

The spatial symmetry of spheres is the highest
and that of lamellae is the lowest. It is natural to speculate that
intermediate phases exist between these two limiting structures.
One typical example is the 2D triangular lattice
formed by cylindrical domains with the azimuthal symmetry (see Fig.~\ref{fig00}C).
For a cylindrical domain with a finite radius $R$ and the infinite height~($L\rtar\infty$),
the surface energy is given by $U_{\txts}= \gamma(2\pi R^2 +2\pi RL) \approx 2 E_0 \alpha {R}^{-1}$
where $E_0=v_\txtd\varepsilon_0=\pi R^2L\varepsilon_0$.
The intra-domain repulsion is derived from the Neumann addition theorem as
\be
U^{(1)}_\txtY&&=2E_0 \left[1-2I_1(R)K_1(R)\right],
\label{eq03_06}
\ee
where $I_n(x)$ and $K_n(x)$
are the modified Bessel functions of the first
and second kinds, respectively.
In this letter we consider the triangular lattice, although
other lattice structures or mixed structures are also possible\cite{leibler}.
We apply the Neumann addition theorem
to calculate the effective repulsion between two domains separated by $r$ as
\be
U^{(2)}_\txtY(r) = 8 E_0 [I_1(R)]^2 K_0(r).
\label{eq03_09}
\ee
The length of the primitive equilateral triangle shown in Fig.~\ref{fig00}C
is related to the volume fraction as $a = (2\pi/\sqrt{3}\phi)^{1/2}R$.
Using this relation, we calculate the lattice energy $U_\txtL$
from the lattice summation of the inter-domain repulsion.
By solving $\partial_R(E_{{\mathrm{tot}}}/E_0)=0$,
we obtain the optimal radius $R_\txtm$ and the energy minimum $E_{{\mathrm{tot}}, \txtm}$.
Similar to other shapes, cylindrical domains are stable for $\alpha<\alpha_c$.

Comparing $E_{{\mathrm{tot}}, \txtm}$ calculated from different shapes
and structures yields the global energy minimum and thus the optimal
ground-state domain morphology. The central result
of this letter is the phase diagram in Fig.~{\ref{fig04},
which describes the shape transformation
between spherical, cylindrical, and lamellar domains.
The ratio $\alpha$ between the surface tension and
repulsion, and the volume fraction $\phi$, are the two control
parameters.
Finite size domains can be stabilized for $\alpha<\alpha_c=1$,
whereas phase separation is observed for $\alpha\ge\alpha_c$.
Domain patterns stabilized in the small attraction
limit ($\alpha\rtar 0$) suggest the possibility of self-assembly
processes in repulsive systems\cite{malescio}.
The basic features of shape transformation
are consistent with symmetry arguments.
In the low density limit~($\phi\rtar 0$),
spheres are preferred at small values of $\alpha$,
whereas lamellae are preferred as the system approaches phase separation.
At larger volume fractions, structures in low dimensions
(2D and 1D) become increasingly more stable. Spheres are
unstable for $\phi>0.19$, while cylinders are unstable for $\phi>0.35$.
At the equal volume fraction of colloidal domains and cavities ($\phi=0.5$),
only the lamellar phase is stable.
The mirror  symmetry between domains and cavities
is used to produce the right half of the phase diagram for $\phi>0.5$.
The cylindrical regime completely separates spherical and lamellar regimes,
demonstrating that 3D spheres undergo a transformation to 1D
lamellae via 2D phases. Although we only compute cylinders, other intermediate shapes
may exist.

\begin{figure}
  \includegraphics[width=7cm]{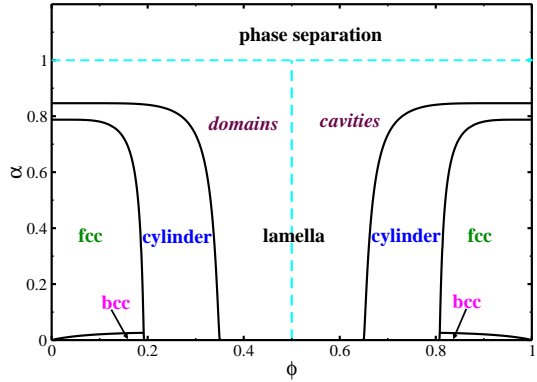}
  \caption{A phase diagram for spherical, cylindrical, and lamellar shapes. The optimal
        structure formed by cavities for $\phi>0.5$ is the mirror of that by colloidal domains at $1-\phi$.}
    \label{fig04}
\end{figure}

In this letter we predict the shape transformation (see Fig.~\ref{fig04})
of colloidal domain patterns in the ground state.
Our study presents a simple and exactly solvable model
system for understanding self-assembling phenomena
based on a pairwise and isotropic potential.
Ground-state domain patterns do not incorporate entropic effects
induced by thermal fluctuations at finite temperatures.
Temperature  effects can be partially included
in the current model by introducing the temperature-dependent surface tension
$\ga(T)$ and domain density $\rho_1(T)$.
At higher temperatures, a more systematic treatment should
involve the calculation of interphase free energies, where the distribution of shapes and
distortion of structures are accounted.
Along this direction, the stability of domains with small distortions,
the liquid-solid transition of particles within spherical clusters,
and the formation of stable clusters at finite temperature are under investigation\cite{caonew}.

This work is supported by the NSF Career Award (Che-0093210)
and the US Army through the Institute of Solidier Nano-technologies at MIT.


\begin{thebibliography}{10}

\bibitem{seul95}
M. Seul and D. Andelman,
\newblock Science {\bf 267}, 476 (1995).


\bibitem{mcconnell01}
H.~M. McConnell and V.~T. Moy,
\newblock J. Phys. Chem. {\bf 92}, 4520 (1988);
H.~M. McConnell,
\newblock Ann. Rev. Phys. Chem. {\bf 42}, 171 (1991);
J.~M. Deutch and F.~E. Low,
\newblock J. Phys. Chem. {\bf 96}, 7097 (1992).


\bibitem{leibler}
L. Leibler,
\newblock Macromolecules {\bf 13}, 1602 (1980);
T. Ohta and K. Kawasaki,
\newblock Macromolecules {\bf 19}, 2621 (1986);
G. H. Fredrickson and E. Helfand,
\newblock J. Chem. Phys. {\bf 87}, 697 (1987);
M. Muthukumar
\newblock Macromolecules {\bf 26}, 5259 (1993);
M. W. Matsen and M. Schick,
\newblock Phys. Rev. Lett. {\bf 72}, 2660 (1994).


\bibitem{blankschtein}
D. Blankschtein, G. M. Thurston, and G. B. Benedek,
\newblock Phys. Rev. Lett. {\bf 54}, 955 (1985);
A. Shiloach and D. Blankschtein,
\newblock Langmuir {\bf 14}, 7166 (1998);
L. Maibaum, A.~R. Dinner, and D. Chandler,
\newblock J. Phys. Chem. B {\bf 108}, 6778 (2004);
S. Tsonchev, G.~C. Schatz, and M.~A. Ranter,
\newblock J. Phys. Chem. B {\bf 108}, 8817 (2004).


\bibitem{gelbart}
R.~P. Sear, S.~W. Chung, G. Markovich, W.~M. Gelbart, and J.~R. Heath,
\newblock Phys. Rev. E {\bf 59}, R6255 (1999).

\bibitem{lipowsky}
R. Lipowsky,
\newblock Nature {\bf 349}, 475 (1991);
G. Ayton and G.~A. Voth,
\newblock Biophys. J. {\bf 83}, 3357 (2002).


\bibitem{widom}
B. Widom, K. A. Dawson, and M. D. Lipkin,
\newblock Physica A {\bf 140}, 26 (1986);
K. A. Dawson,
\newblock Phys. Rev. A {\bf 36}, 3383 (1987).

\bibitem{wolynes00}
J. Schmalian and P.~G. Wolynes,
\newblock Phys. Rev. Lett. {\bf 85}, 836 (2000);
S.~W. Wu, H. Westfahl Jr., J. Schmalian, and P.~G. Wolynes,
\newblock Chem. Phys. Lett. {\bf 359}, 1 (2002).

\bibitem{Verwey}
E.~J.~W. Verwey and J.~TH.~G. Overbeek,
\newblock {\em Theory of the Stability of Lyophobic Colloids},
\newblock (Elsevier, Amsterdam, 1948).

\bibitem{Wu}
C. Wu and S.~H. Chen,
\newblock J. Chem. Phys. {\bf 87}, 6199 (1987).

\bibitem{sciortino}
F. Sciortino, S. Mossa, E. Zaccarelli, and P. Tartaglia,
\newblock Phys. Rev. Lett. {\bf 93}, 055701 (2004);
S. Mossa, F. Sciortino, P. Tartaglia, and E. Zaccarelli,
\newblock Langmuir {\bf 20}, 10756 (2004);
F. Sciortino, P. Tartaglia, and E. Zaccarelli,
\newblock arXiv:cond-mat/0505453.

\bibitem{caowu042}
J.~L. Wu, Y. Liu, W.~R. Chen, J.~S. Cao, and S.~H. Chen,
\newblock Phys. Rev. E {\bf 70}, 050401 (2004).


\bibitem{fcc_bcc}
P. M. Chaikin, P. Pincus, S. Alexander, and D. Hone,
\newblock J. Colloid and Inter. Sci. {\bf 89}, 555 (1982);
K. Kremer, M. O. Robbins, and G. S. Grest,
\newblock Phys. Rev. Lett. {\bf 57}, 2694, (1986).

\bibitem{malescio}
G. Malescio and G. Pellicane,
\newblock Nat. Mater. {\bf 2}, 97 (2003).

\bibitem{caonew}
J. L. Wu and J. S. Cao,
\newblock accepted by J. Phys. Chem. B;
A. Zhukov and J.~S. Cao,
\newblock In preparation.


\end{thebibliography}
\end{document}